\documentclass[aps,preprint]{revtex4}

\begin{document}
\bibliographystyle{prsty}



\title{Snapshots on Vortex Dynamics}

\author{ P. Ao }
\address{Dept's of ME and Phys.,
         Univ. of Washington, Seattle, WA 98195, USA }
\date{ April 7, 2005 }


\begin{abstract}
 Salient features of vortex dynamics in super media are summarized.
Recent examples are: the demonstration of prominent role of
topology in vortex dynamics; the solution to the Hall anomaly
which once bothered Bardeen, de Gennes and many others; the
unified microscopic treatment of both transverse and frictional
forces on moving vortex.  The fundamental dynamical equation of
vortex matter can now be casted into the elegant form of quantum
dissipative dynamics of Leggett. Together with the
Kosterlitz-Thouless transition, we have finally reached a coherent
picture on both thermodynamic and dynamical roles played by
vortices. The key historical progresses are discussed with a
broader perspective, to move into the post high $T_c$
superconductor era, the quantum era.

References mentioned in the text and given at the end, though very
incomplete, along with a list of a few outstanding open problems,
may provide a reader a useful guidance and an interesting
perspective.

\noindent
{\bf Puzzle}:  \\
By 1999 Kopnin and Vinokur reached the conclusion that the
anomalous Hall effect can be compatible with the Magnus force,
though the present author reached the same conclusion 4 years
earlier but was not cited by them. By 2001 Blatter and Geshkenbein
and their coworkers reached the conclusion that in discussion of
vortex interference effect only the vortex velocity part of Magnus
force is needed, though again same conclusion was reached by the
presented author 5 years earlier but was not cited by them. It is,
however, very comforting that those important physics have been
explored by very different groups of able physicists.

The puzzle here is not on their inability to cite relevant prior
works, for an analysis of such behaviors the readers are referred
to the Kirby-Houle article in Nov. (2004) Physics Today. Instead,
the puzzle is on their ability to maintain (Kopnin, 2001; Blatter
and Geshkenbein, 2003) that there is reduced and/or sign-reversed
transverse force without giving any discussion on the
contradiction to their own as well as other related works.



\end{abstract}



\maketitle

\section{ Fundamental Vortex Dynamics Equation}

\subsection{Fundamental equation for vortex matter}

All the fundamental features regarding to vortex dynamics are
already present in two dimensions. The generalization to three
dimensions is straightforward. After long and strenuous efforts by
Ao, Geller, Niu, Rhee, Tang, Thouless, Wexler, Zhu, and many
others, elegant formulation of vortex dynamics and its proper
physical understanding have apparently been reached.

In two dimensions, the fundamental equation of motion for a vortex
reads:
\begin{equation}
 m_v  \frac{d^2{\bf  r}(t) }{d t^2}
   = - \nabla V({\bf r}(t)) -  2\pi \hbar q_v \rho_s({\bf r}) \frac{d {\bf r}(t)}{d t}
             \times \hat{z}
     - \int_{-\infty}^{t} dt' \; \alpha (t-t') \frac{ d {\bf r}(t')}{d t'} + \xi(t)
\end{equation}
with the correlation function
\begin{equation}
   \alpha(t)
     = \frac{2}{\pi} \int_0^{\infty} d\omega  \frac{J(\omega)}{\omega} \; \cos(\omega t )
\end{equation}
and the spectral function
\begin{equation}
    J(\omega) = \eta \omega ^{s} \exp\left( -\frac{\omega}{\omega_c} \right)
\end{equation}

Here ${\bf r}$ is the vortex position vector, $\hat{z}$ is the
unit vector perpendicular to the plane; $\rho_s({\bf r})$ is the
superfluid density at the vortex, the Planck constant $\hbar$, and
the vorticity $q_v$ which is an integer. The cutoff frequency
$\omega_c$ will be chosen to be larger than any characteristic
frequency in the problem.  All other terms, the potential $V$, the
transverse force, the frictional force, and the noise, with be
explained below.

Ao and Thouless (1993, \cite{at}) and Thouless, Ao, Niu
(1996,\cite{tan}) are two most important theoretical progresses
during 1990's in the understanding of vortex dynamics. They have
been served as the light houses in the navigation through the
vortex dynamics rough water. To my knowledge Ao and Zhu (Ao and
Zhu, 1999 \cite{az}) is the best place to get into vortex dynamics
in the formulation of Eq.(1) from a microscopic point of view. The
general physics behind such equation, system plus environment, in
the context of dissipative quantum dynamics can be found, for
example, in Leggett (1992) or in Feynman and Vernon (Ann. Phys.
1963).

The well-known linear friction case is a special case of Eq.(1):
\begin{equation}
 m_v  \frac{d^2{\bf  r}(t) }{d t^2}
  = - \nabla V({\bf r}(t)) -  2\pi \hbar q_v \rho_s({\bf r}) \frac{d {\bf r}(t)}{d t}
             \times \hat{z}
    - \eta  \frac{ d {\bf r}(t) }{d t} + \xi(t)
\end{equation}
It is the Ohmic case that $s=1$ in the spectral function, the
cutoff frequency goes to infinite ($\omega_c \rightarrow \infty$),
the Planck constant is zero ($\hbar = 0$) or the high temperature
limit.

\subsection{Nonlinear Schr\"{o}dinger equation}

The most physically transparent and mathematically elegant way to
derive Eq.(1) is to from the nonlinear Schr\"{o}dinger equation
(NLSE): Every term in Eq.(2) can be obtained from NLSE. For
topological quantities, such as the Magnus force, NLSE is
sufficient. But some contributions to friction, such as the core
and extended quasi-particle states, have to be calculated from
microscopic theories. The advantage of starting from NLSE is that
the connection to hydrodynamics and to quantization is direct. For
example, for the neutral case, $s=2$, a superohmic case according
to Leggett's formulation of dissipative dynamics, while for the
charged case, $s=\infty$, that is, there is a gap, the Coulomb
gap, in the elementary excitations.

The NSLE in both neutral and charged cases can be derived from
microscopic theories. The simple form of NLSE reads:
\begin{equation}
 i\hbar \frac{ \partial \psi({\bf r},t) }{\partial t}
  = - \frac{\hbar^2}{2m}\nabla^2 \psi({\bf r},t) - \mu_0 \psi({\bf r},t)
    + U_0 |\psi({\bf r},t)|^2 \psi({\bf r},t)
\end{equation}
With $\rho_s({\bf r},t) = |\psi({\bf r},t)|^2$ the superfluid
density at time $t$ and position ${\bf r}$, $m$ the effective mass
of the Cooper pairs or bosons, $\mu_0$ the chemical potential
determined the mean superfluid density, and $U_0$ the effective
strength of the short range repulsive interaction.  Phonons and
vortices are automatically included in this formulation. The
macroscopic slow dynamics of NLSE is completely exhausted by
dynamics of phonons and vortices.

The consistence of NLSE with microscopic theory for superfluid
Helium was pointed out by Demircan, Ao, Niu (1996). Such
connection was already implicitly known to Feynman and to
Anderson. NLSE can also be obtained from Kohn's density functional
approach.

Josephson relation can follows directly from this NLSE, as shown
by Feynman.

The short length scale in NLSE is the healing length or coherence
length determined by $U_0$, the important short length scale for a
macroscopic description. The introduction of coupling to
electromagnetic field is straightforward: the standard minimum
coupling. In this case another length scale, the London
penetration depth connected to the superfluid density, enters into
the description. Therefore, the even the usual two types of
superconductors, type I and type II, can be effectively described
by NLSE.

There should be no confusion of NLSE with Gross-Pitaevskii
equation (GPE): GPE is about the off-diagonal part, the condensate
part (first clearly conceived by London in 1948), which is highly
sensitive to the strength of the interaction among the fluid
particles, bosons or fermions. Instead, NLSE is about the
superfluid density, which is always the total fluid density (for
simple fluid) at zero temperature regardless of the interaction
strength. For example, for a strongly interacting bosons, such as
He II, the condensate can be a small fraction of the superfluid
density at zero temperature.

The current description of BEC at zero temperature is in a happy
situation: at zero temperature, the GPE and NLSE are almost
identical, because the interaction is weak. Nevertheless, for the
discussion of vortex dynamics, physical it is the NLSE not GPE
which one should use and keep in mind to avoid confusions.

The derivation of NLSE in superconductor from BCS theory was given
by Aitchison, Ao, Thouless, Zhu (1995).  There has been a
consideration amount of confusion between NLSE and time dependent
Ginzburg-Landau equation (TDGL) till these days. TDGL is
essentially a GPE equation (and vice versa) in fermionic
superfluid: strictly it is about the condensate fraction, not the
superfluid density. Again, at zero temperature, there is no simple
relation between the gap function in TDGL and the superfluid
function $\psi$. For example, the gap can be exponentially small
but the superfluid density will be the total free electron
density. One should not be surprised that TDGL can take a complete
different form from that of NLSE. Nevertheless, we have another
happy situation in superconductors: near transition temperature
$T_c$ the super fluid wavefunction $\psi$ was shown by Gorkov to
be proportional to the gap function in TDGL.

In the present of weak disorder in superconductors, NLSE will
retain its form of Eq.(5), with the same superfluid density
implied by Anderson's dirty superconductors theorem and justified
by Green's function approach by many others, but with a different
effective mass known to Pippard.

\subsection{Vortex mass $m_v$}

Vortex mass is perhaps the first example of the acquiring mass
from the environment, discussed more than 100 years ago. It is the
first example of the renormalization of mass. However, it is also
interesting to point out that it is perhaps the least
experimentally tested quantity in this category.

It is effectively the mass of the fluid excluded by the vortex
core, for the ideal incompressible fluid.

This mass can be calculated. The hydrodynamics case can be found
in H Lamb's classical book. The superconductor case can be found,
for example, in   Han et al (Han, Kim, Kim, Ao, 2005).

In the slow dynamics limit the left hand side of Eq.(1) is a
higher order contribution to dynamics. It may be negligible. Then
the dynamics would be dominated by the Lorentz force like
transverse force and/or the correlation function which contains
the dissipation.

This may explain the difficulty in experimental measurement of
vortex mass: For slow dynamics, it's contribution is of higher
order, and for a relative fast dynamics, the dissipative effect
becomes large. Hence, a very precise measurement should be needed
in order to have reliable number on the vortex mass. This implies
that a different type of experimental design, other than those to
measure the potential, transverse force, and friction, is needed.

\subsection{Vortex potential $V({\bf r})$ and its gradient}

The potential includes all the contributions which are not
dependent on the vortex velocity. More precisely, all the
positional dependence in this term is instantaneous.

It contains a term coming from the fluid velocity generated by
others vortices, including those from the image of the vortex
under consideration. It's gradient has the form, the superfluid
velocity part of the Magnus force:
\begin{equation}
  F_{Magnus, {\bf v}_s}
   =  2\pi \hbar q_v \rho_s({\bf r}) {\bf v}_s({\bf r})  \times \hat{z}
\end{equation}
If there is no other terms such as pinning in $\nabla V$, trapping
potential in in BEC, and no frictional force and noise, this term
together with the transverse force is the known Magnus force in
fluid dynamics. It makes the vortex moving along the superfluid
flow stream line.

Some famous results have obtained from this term which describes
the vortex-vortex interaction:

{\bf a)} Critical velocity. Feynman (1954), Anderson ({\it e.g.},
Basic notions of condensed matter physics, 1984), Leggett (Physica
Fennica, 1973).  The meaning of critical velocity is firmly placed
on the topology, not of Landau critical velocity type of
quasiparticle with no topology.

There is, however, another happy situation. In many cases the
numerical values of critical velocity due to Landau and due to
topological consideration are the same, or, very close to each
other, though in general it has been shown by Anderson and by
Leggett that there is no relation between them.

{\bf b)} Abrikosov vortex lattice:  vortex form lattices.  \\
This force leads to logarithmic interaction in neutral case and a
short range (on the scale of London penetration depth) in the
charged case. An equilibrium lattice structure almost follows
immediate this way.

{\bf c)} Kosterlitz-Thouless transition:
  the unbinding of vortex-antivortex  pairs. \\
This transition is extremely important in the understanding of the
topological stability of condensed phase, and resulting in the
name of Kosterlitz-Thouless-Halperin-Nelson-Young transition.

{\bf d)} Quark confinement and asymptotic freedom. \\
Kosterlitz-Thouless transition is also an elementary (2D)
illustration of the quark confinement (The phases of quantum
chromodynamics : from confinement to extreme environments. JB
Kogut and MA Stephanov. Cambridge University Press, 2004).

\subsection{Transverse force: the vortex velocity part of Magnus force}

This transverse force is the second term at the right hand of
Eq.(1), identical in form to the Lorentz force:
\begin{equation}
 F_{Magnus, d{\bf r}/dt }
   = - 2\pi \hbar q_v \rho_s({\bf r}) \frac{d {\bf r}(t) }{d t} \times \hat{z}
\end{equation}

It's derivation from microscopic theory   (Ao and Thouless, 1993)
is one of the nontrivial applications of Berry phase to obtain
important physical results. The topological structure of a vortex
had been discussed by London (1948), Onsager (1949), and Feynman
(1954).

The first macroscopic derivation of Eq.(7) was given by Nozieres
and Vinen (1966). See also Fetter, PR 163, 1967.

It is another expression for the Josephson-Anderson relation.
Anderson, RMP, 1966; ME Fisher and Langer, PRL, 1968.

The full detailed microscopic derivation in superconductors was
given by Ao and Zhu (Ao and Zhu, 1999), including both the
contributions from the vortex core and extended states, as well as
in both clean and dirty limits. The feasibility of such derivation
is guaranteed by the Anderson's dirty superconductor theorem.

This force has rich physics consequences in addition to the
Josephson-Anderson relation, for example:

{\bf a)} turbulence (Onsager, 1949);

{\bf b)} anomalous Hall effect in superconductor (Ao, 1995; Kopnin
and Vinokur, 1999);

{\bf c)} vortex interference (van Wees, 1990; MPA Fisher, 1991; Ao
and Zhu, 1995);

{\bf d)} quantum Hall effect in Josephson junction arrays (Zhu,
Tan, and Ao, 1996);

{\bf e)} vortex processing in BEC (Lundh and Ao, 2000)

{\bf f)} interference effect (Ivanov, Ioffe, Geshkenbein, Blatter,
2001)

Experimental evidences are numerous to support the above
theoretical proposals. It is clear that by 1999 theoretically
there exists an agreement that the anomalous Hall effect is
consistent with the transverse force as given by Eq.(7).

It is also clear by 2001 that in order to consider the transverse
effect on vortex motion in Josephson junctions arrays, Eq.(7) is
the only transverse force responsible for various quantum effects.
No other transverse effects introduced by various authors are
needed in such discussions.

\subsection{Frictional force
  $- \int_{-\infty}^{t} dt' \; \alpha (t-t') \frac{ d {\bf r}(t')}{d t'}$ }

For $ 2 \geq s \geq 0 $, if one perform the usual effective energy
calculation with constant vortex velocity,
infinite vortex mass correction will be resulted: \\
For $s=2$, the effective mass correction will diverge
algorithmically with systems size, a fact elegantly discussed by
Duan and Leggett (1995) and confirmed by Niu, Ao, Thouless (1996)
via a dynamical and many-body wavefunction consideration.

From the microscopic derivation, one contribution to $s=1$ was
first found by Bardeen and Stephen (1965) from the vortex core in
the dirty limit. $s=1$ was also found by Ao and Zhu (1999) from
the extended state contribution. Such contributions are the Ohmic
type:
\begin{equation}
  - \int_{-\infty}^{t} dt' \; \alpha (t-t')
  \frac{ d {\bf r}(t')}{d t'} \rightarrow
    - \; \eta \; \frac{ d {\bf r}(t)}{d t}
\end{equation}

For $s < 0$ the system is thermodynamically unstable. For $s>2$,
vortex mass renormalization due to $\alpha(t)$ and $\xi(t)$ is
finite. The diverging mass encountered here ($ 2 \geq s \geq 0$)
is closely related to those diverging quantities in non-Fermi
liquid theory.

The regime $ s \geq 0 $ also makes the adiabatic consideration of
vortex motion possible, though in the regime $ 2 \geq s \geq 0 $ a
strict Landau quasiparticle type picture (finite effective mass
etc) is not valid.

The very existence of this friction force implies, in addition to
the effective mass, that vortices can be independent variables: it
will not necessary move along the superfluid flow stream line, and
can cut through the streamlines. Thus, the vortex motion can
generate dissipation, even when the fluid is ``super", a common
knowledge now in superfluid and superconductors, after several
Nobel prizes.

\subsection{Noise $\xi$ }

The noise is related to the friction by the
fluctuation-dissipation theorem, derivable from microscopic
theories:
\begin{equation}
  \langle \xi(t) \xi^{\tau}(t') \rangle
    = \frac{\hbar}{\pi} \int_0^{\infty} d\omega J(\omega)
      \coth\left(\frac{\hbar\omega}{2k_B T } \right) \cos(\omega(t-t'))
\end{equation}
and $\langle \xi \rangle = 0$. Here superscript $\tau$ denotes the
transpose. For simplicity we have assumed the friction matrix to
be a constant matrix.  No anisotropic frictional effect will be
considered here.

Such an expression can be derived either starting from NLSE or
from microscopic theories: we already mentioned that the
vortex-phonon interaction corresponds to $s=2$ and core and
extended states contributions correspond to $s=1$.

In the zero $\hbar$  or high temperature limit, we have for $s=1$,
\begin{equation}
  \langle \xi(t) \xi^{\tau}(t') \rangle = 2 k_B T \; \eta \; \delta (t-t')
\end{equation}
This corresponds to Eq.(4).

\section{ Some High and Low Points  }

Here are snapshots on the progress in vortex dynamics, emphasizing
on superfluids and superconductors.

\subsection{ Pre-high $T_c$ superconductor era ( $<$ 1989 ) }

Vortices were not in Landau's original formulation of two fluid
model of Helium II. In fact, Landau initially opposed the
existence of the vortices. This ``absence of vorticity" might be
the origin of confusing on vortex dynamics from the theoretical
side.

{\bf  1965, Bardeen and Stephen.}
   Microscopic calculation of vortex friction on core contribution in
the dirty limit \cite{bs}. An elegant paper perhaps has not been
widely read, though widely cited. The misunderstanding on the
origin of friction still exists.

{\bf  1966, Nozieres and Vinen. }
   Macroscopic derivation of Magnus force \cite{nv}. Very insightful paper.
   A. Fetter's 1967 PR paper is also helpful.

{\bf 1976, Noto, Shinzawa, Muto.}
   Summarizing the Hall anomaly experiments in superconductors: the Hall
effect is usually small and often change signs, in an apparent
contradiction to the transverse force as given by Eq.(7) if using
the independent vortex dynamics model to calculate the Hall
effect.

   Similar effect has been observed in superfluids.

   This ``anomalous" effect might be the origin of confusing on
   vortex dynamics from the experimental side.

{\bf 1976, Kopnin and Kratsov.}
   In response to the small Hall
effect in the mixed state, relaxation time approximation was
conceived by Kopnin and Kratsov to derive the core friction
contribution with vanishing small transverse force in the dirty
limit.

In the hindsight, this approximation is not applicable in this
case. The physical and mathematical reasons for such a invalid
approximation have been discussed at least since 1940's. In
particular, R. Kubo had extensively discussed such approximation
(Statistical physics, M. Toda, R. Kubo, and N. Saito, v.1 and 2,
second edition, 1992). See also Zubarev of Bogoliubov school
(Nonequilibrium statistical thermodynamics, D. N. Zubarev, 1974)
for appropriate time scales in the problem.

{\bf 1976, Sonin.}
   Approximated calculation of additional transverse force due to phonons.
There is no clear interpretation of the additional force by Sonin,
such as whether add or subtract to the transverse force as given
in Eq.(7). The direct contradiction of such result with Vinen's
experiment has never been discussed.

Theoretically, NLSE gives a complete description at zero
temperature: the superfluid density is the total fluid density,
there are vortices and phonons, and vortex and phonons interact.
There is no additional contribution from phonons to the super
(total) fluid density.

\subsection{ High $T_c$ superconductor era ( $>$ 1989 ) }

{\bf 1993, Ao and Thouless.}
   Berry phase derivation of the Magnus force based on topology
and symmetry of many-body wavefunction \cite{at}. A nontrivial
application of Berry's method. The topological aspect of vortex
dynamics was emphasized.

{\bf 1993, Volovik.}
   The absence of transverse force in the dirty
limit was interpreted as the cancellation between the topological
contribution from the core stated, the spectral flow, and the
topological effect of Berry phase.

This is an erroneous conclusion. The fact is that there are two
equivalent ways to compute the transverse force on a moving
vortex: one from core regime and one away from core. They are
equal according to Stokes theorem.

The spectral flow is independent of the impurity because of its
topological nature, not something continuously tunable by a non
topological parameters such as the relaxation time.

Hence, two mistakes would not make it right.

{\bf 1995, Kopnin and Lapatin; van Otterlo, Feigelman,
Geshkenbein, Blatter.}
   Repeating the relaxation time approximation to Helium 3 by the
first group author and extended to  superconductor with under path
integral formulation by the second group authors.  Confirmed their
old conclusion that there is no transverse force in the dirty
limit.

Again, the mistake is the invalidity of the relaxation time
approximation.

{\bf 1995, Feigelman, Geshkenbein, Larkin, Vinokur.}
   Trying to demonstration an additional Berry phase term from the
vortex core to cancel the Berry phase computed by Ao and Thouless
(1993).

   Their calculation is in clear violation of the basic requirement from
quantum mechanics: at the phase singularity the amplitude of the
wave function must be zero. Hence, there is no additional Berry
phase term as they claimed.

   However, since their result apparently reproduced what obtained
by Kopnin and Krastsov, by Volovik, and the results of their other
collaborators based on erroneous approximation schemes, they
believe their cancellation should be right. Hence, they found that
not only the transverse force is usually small, it occasionally
changes signs, controlled by relaxation time, etc.

{\bf 1995, Ao.}
   First explicit proposal that the large transverse force is
consistent with Hall anomaly if vortex many-body and pinning effects are
considered \cite{ao95}. Several quantitative predictions were made
here.

   The main conclusion is that, the anomalous Hall effect in the mixed state,
the small Hall angle and sign change, can be explained by the
universal Magnus force derived from the Berry phase. This may not
be a surprising result for people familiar with the Hall effect in
semiconductors: there we see small and zero Hall angle, size
changes, etc, and they are all consistent with the universal
Lorentz force.

{\bf 1995, Ao and Zhu.}
  Vortex interference by controlling the number of particles in the
superfluid enclosed by the vortex trajectory loop \cite{az95}.

  Since the transverse force is similar to the Lorentz force, this
is just another form of Aharonov-Bohm effect for vortices.

{\bf 1996, Thouless, Ao, and Niu.}
  Extension of Berry phase formulation to include the friction \cite{tan}.
No relaxation time approximation is needed. But a proper
thermodynamical limit is required: the dissipative energy has been
carried out of the system, preferably to infinite in an explicit
manner.

 This is a nontrivial extension of Berry's method.  Mistakes
have often been committed in such an extension. The discussions of
R. Kubo mentioned above as well as those by Zubarev are useful
here for a better understanding of physics.

{\bf 1996 Zhu, Tan, and Ao.}
  Quantum Hall effect in Josephson junction arrays from the view
of vortices \cite{zta}.

{\bf 1997, Sonin.}
   Same approximated calculation as his 1976 was repeated.  It is clear that
even within such approximation, the linear correction term wanted
by Sonin cannot be rigorously obtained. But this mathematical
inconsistency was completely ignored by Sonin in order to generate
result he wanted.

The present of phonons and the total superfluid density is equal
to the total fluid density at zero temperature implied in NLSE
(Eq.(5)) clearly suggests Sonin's concept here is completely
wrong.

By a careful analysis, it should be concluded that what Sonin
discussed was a different phenomena other than what he thought.
After all it cannot produce what he wanted in a mathematically
consistent manner.

{\bf 1997, Zhu, Brandstrom, and Sundqvist.}
  First direct confirmation of transverse force on vortex in
superconductor \cite{zhu}. This elegant experiment was done  in
the tradition of directly measuring the Lorentz force for electron
in the magnetic field (1890's) and vortices in superfluid
(1960's).

It is very surprising that despite over 30 years controversies on
the transverse force, this is the only systematic experiment to
directly measure the force in superconductors.

{\bf 1999, Ao and Zhu.}
  Detailed and microscopic implementation \cite{az} of framework
developed in 1996 by Thouless, Ao, Niu.

  The results of Bardeen and Stephen and of Nozieres and Vinen
were unified and extended. Detailed calculation showed how to
obtain the vortex friction without the relaxation time
approximation, consistent with what Bardeen and Stephen did.

   It is interesting to note that there is no controversy at all
on Eq.(6) (Wexler, PRL, 1997). From the macroscopic hydrodynamical
point of view Eq.(6) and (7) are just the two sides of same coin.

  Thermodynamically, it was demonstrated by Ao and Zhu that the
reduction of total transverse from Eq.(7) as fiercely argued by
Blatter, Feigelman, Geshkenbein, Kopnin, Larkin, Vinokur, Volovik,
and others (their conclusions are all based on uncontrolled
approximations) would lead to the violation of the second law of
thermodynamics.

{\ }

To summarize what done by Ao and Zhu, the invalidity of relaxation
time approximation was carefully considered from both critical and
constructive points of views, from both macroscopic and
microscopic points of views:

{\bf a)} An elementary kinetic model was devised in Ao and Zhu
(PRB 1999), adapted from Kubo and others, to show how the seemly
simple use of relaxation time approximation lead to wrong result.

The essence of the demonstration is that, in the calculation of
transport coefficients, there are usually two different starting
points for systematic approximation, though rigorously they are
equivalent in the linear regime. The first one is to treat the
force as perturbation and calculate the response velocity
(current):
\[
 {\bf small \; force} {\ } \Longrightarrow  {\ } {\bf velocity}
   {\ } , {\ }  (I) \; .
\]
In this case there is usually a well-defined expression for the
force to begin with, and the velocity-velocity (current-current)
correlation is the one subjected to systematic approximation. The
well-known example in this category is the conductivity. The
relaxation time approximation is usually OK, and one can simply
start from a typical kinetic equations such as the Boltzmann
equation.

The second starting point is to treat the current, or velocity, as
the perturbation and calculate the response force.
\[
 {\bf small \; velocity} {\ } \Longrightarrow {\ } {\bf force}
     {\ } , {\ } (II) \; .
\]
This method also has other names, such as the force-balance
equation. It is the force-force correlation subjected to
systematic approximation. The well-known example here is the
computation of resistivity. Unfortunately, the usual relaxation
time approximation is problematic here, documented over past 50
years in literature.

In the case of derivation of vortex dynamics microscopically, we
do NOT know that form of the force on a moving vortex at
beginning: It is precisely this force needed to be found out.
Hence, we cannot use the usual approach of starting from using the
force as perturbation. We are compelled to deal with the second
one, using the vortex velocity as the perturbation. This is what
has been used by all of us: Thouless {\it et al}, Kopnin {\it et
al}, van Otterlo {\it et al}, and so on.  As it is known in
literature, one should avoid the problematic relaxation time
approximation in this case. But Kopnin {\it et al}, van Otterlo
{\it et al} etc have not.

A sophisticated and clear demonstration can be found in Kubo's
book as well as in the book of Zuburev, mentioned above: there are
several time scales important at the microscopic level, but not
apparent at the macroscopic level.  One has to be carefully on the
limiting procedure. Kubo himself had complained about the blind
and wrong use of relaxation time approximation in transport
problems, which appears periodically in literature.

One may put it in following way:  It is the relaxation time
approximation which needs to be justified here: Boltzmann
recognized this long long time ago. It arises from the interaction
between different parts of the whole system. Here, in the context
of vortex dynamics, the friction of vortex directly comes from the
interaction of vortex with the quasiparticles, and can and have
been calculated without the relaxation time approximation. If one
wants an expert understanding of this issue, Leggett's formulation
of dissipative quantum dynamics and Kubo's book are among the must
readings. We may summarize what has been known in transport theory
in the following table:

{\ }

\noindent
\begin{tabular}{|c|c|c|}
  \hline
   & (I): force as cause & (II): velocity as cause \\
  \hline
  physical applications: & deriving Drude formula;
       & deriving vortex dynamics; \\
    examples   & application of Boltzmann equation;
       & Berry phase calculation; \\
       & velocity-velocity correlation
       & force-force correlation \\
  \hline
  question of validity of & usually OK & usually problematic \\
  relaxation time approximation &  &   \\
  \hline
  references & condensed matter physics books;
       & Kubo's book; \\
       & Green functions approaches  & Zubarev's book  \\
  \hline
\end{tabular}

{\ }

 {\bf b)} A thermodynamical demonstration was also devise to
show that the change in superfluid kinetic energy must come from
the transverse force on the moving vortex, since the entropy of
superfluid is zero. Any reduction of the magnitude of this
transverse force, as would be the case for the present relaxation
time approximation, will violate the second law of thermodynamics.
This gives a thermodynamical reason to abandon the relaxation time
approximation in this case.

{\bf c)} A full microscopic derivation of the transverse was
provided in Ao and Zhu (PR B 1999). It was a detailed
implementation of the formulation developed by Thouless, Ao, and
Niu (PRL, 1996). It is very important to point out that this
microscopic formulation is similar to what used by Kopin et al, by
van Otterlo et al, and by many others. The only major difference
is the absence of relaxation time approximation in the context of
vortex dynamics in Ao and Zhu.

It was found that there are many contributions to the vortex
friction: core states, extended states, etc. The contribution of
core states is due to the mixing of core levels by impurity
scattering under an appropriate time scale. This mixing
contribution to friction has been known for a long time,
reminiscent to Thouelss energy, at least since 1980's, and has
been made very clear and explicit in the recent study of chaotic
contribution to friction.

{\ }

{\bf 1999, Kopnin and Vinokur.}
  The compactibility of large transverse force in Eq.(7) with Hall anomaly
was argued \cite{kv}, though no citation to Ao and/or Thouless.

Experiment of Zhu et al on transverse force was cited by Kopnin
and Vinokur.

It is very comforting that the same result obtained by Ao four
years earlier on Hall anomaly was reached by a very different
group of able physicists.

At least as late as in 1999, one may be able to conclude that the
Hall anomaly is compatible with the large transverse force.

{\bf 2001, Ivaon, Ioffe, Geshkenbein, Blatter.}
  Vortex interference and its effect in Josephson junction arrays
were discussed solely based on Eq.(7) \cite{iigb}, with no
citation to Ao and/or Thouless.  There is however no presence of
other transverse forces discussed by Geshkenbein, Blatter, and
others in their earlier works. Indeed, such additional transverse
forces are not needed, either.

Even if those authors still believe in the existence of other
extra transverse forces,from the professional point of view they
should state that either the extra forces are not needed or they
do not exist in this situation. Those authors should also state
that there is an alternative theory by Ao and Thouless, and by
others that no such extra forces at all.

Again, it is very comforting same physics explored a few years
earlier was done by a very different group of able physicists. NO
other transverse forces, such as discussed by kopnin et al, by van
oterrlo et al, by Feigelman et al, by Volovik, etc, on a moving
vortex is needed.

{\ }

{\bf Summary:} We may be able to conclude that the controversy on
the transverse force of Eq.(7) may be finally behind us. There is
no reduction from Eq.(7).

\subsection{ Post high $T_c$ superconductor era ( $>$ 1998 ) }

The post high $T_c$ superconductors era study is marked by
Bose-Einstein condensation, topological controlled quantum
computation, quantum turbulence, etc. It is the explicitly
exploration of quantum behaviors of vortices, hence the quantum
era. It looks that we have finally, by 1999 if earlier or by 2001
if later, reached a firm understanding on vortex dynamics, in the
form given by Eq.(1), and in particular the transverse force in
the form of Eq.(7).

However, it appears that rest of us are all too naive.

{\bf 2001, Kopnin.}
  In 2001 Kopnin still presented the situation as based on his erroneous
theory of vanishing transverse force in the dirty limit \cite{v2}.
There is no citation to Ao and/or Thouless.

It appears that he has never consulted the discussion by R. Kubo
on the invalidity of relaxation time approximation in certain
approach, including his case. No discussion on his violation of
second law of thermodynamics as demonstrated by Ao and Zhu in
1999.

{\bf 2003, Blatter and Geshkenbein.}
  In 2003 Blatter and
Geshkenbein misled the community by systematic suppressing the
literature on the existence of transverse force and by emphasizing
their works on the vanishing of transverse force in the dirty
limit \cite{v1}.

{\ }

There are some questions here before moving on.

Should the new era start from erroneous results which has been
demonstrated? Let's put aside the question on Kopnin, Vinokur,
Blatter, Geshkenbein systematic omission to relevant prior work.
What is the logic behind Kopnin as well as Blatter and Geshkenbein
that when they need the transverse force, it is there, and when
they don't, they simply announce that it would not exist?

\section{ Future }

First of all, it seems that the research community deserves
explanations (none so far) from Kopnin, Vinokur, Blatter,
Geshkenbein, and their collaborators, on their inconsistent
behaviors regarding to the use of Eq.(7), the transverse force.
This is science. Researchers deserve honest answers.

It is difficult to predict what will be the exciting results
coming out of BEC, quantum computation, and quantum turbulence and
other related fields. Here I would rather focus on the unsolved
problems along the more traditional line. Their solutions will
undoubtedly help us understand other problems.

 {\bf treatment of boundary layer.}
   On the phenomenological level
of two fluid model, it would be nice to further extend the result
obtained by Thouless, Geller, Vinen, Fortin, Rhee (2001) (Also,
Rhee, PhD Thesis, 2003, University of Washington). The approach
will very likely be based on a method on the treatment of boundary
layer. This will not only gain an understanding on vortex
dynamics, it may also lead to a new insight on boundary layer
problem in general. However, this may be a rather hard problem.

{\bf microscopic derivation of fundamental equation in bosonic
superfluid.}
   Though mathematical framework to calculate
the transverse force and friction on a moving vortex has been set
up by Thouless et al (Thouless, Ao, Niu, 1996), and such a
calculation has been performed for superconductors (Ao and Zhu,
1999), strangely enough there is no full range calculation yet for
bosonic superfluid even based on Bogoliubov theory. There is an
apparent difficulty due to the existence of both infrared and
ultraviolet divergences built into the conventional approximation.
In calculation of the friction and transverse force, a consistent
consideration of all degrees of freedom is needed, because of the
topology of vortex.

Hence, one needs to develop a consistent microscopic superfluid
theory at finite temperatures based on Bogoliubov formulation. The
rest of calculation on vortex dynamics would strongly resemble
what has been done by Ao and Zhu (1999). A major difference may be
that there is now no localized core states. I expect to see a
progress soon along this direction, and I would be delighted to
hear it as soon as possible ({\bf E-mail:}
aoping@u.washington.edu).

{\bf phonons and quasiparticles.}
   For fermionic superfluid, the different between phonons and
quasiparticles who carry super current is as clear as blue sky.
This is not so in bonosic superfluid. We know that phonons exist
in both super and normal phases, but supercurrent exists only in
one phase. Even in super phase, the receiver of Packard can get
phonons in Helium II, not supercurrent. A clear understanding of
this issue may deepen our understanding of the microscopic theory
of bosonic superfluid, and may help understand some issues in
quantum tuubulence, too.

{\bf measurement of transverse force and friction.}
   There is no doubt that more precise as well as further
measurements on both transverse force Eq.(7) and friction Eq.(8)
are needed, for both bosonic and fermionic superfluids. Such
experiments really require courage and talent: Courage to initiate
experiments and talent to design good experiments. This has
already been demonstrated \cite{vinen,packard,zhu}. It is
surprising that how little has been done experimentally to test
the fundamental vortex equation.

{\ }

Though it is impossible in this short note to give a complete
reference list, I do hope that the works mentioned in the text
should give a reader a useful guidance to literature with an
interesting perspective. From my own point of view, a good entry
point to have an overview may be Ao and Zhu (1999) \cite{az}. Some
of additional relevant comments on the theoretical side may be
found in Ref.[\cite{ao03,ao04}].

\end{document}